# Prompt Injection as an Emerging Threat: Evaluating the Resilience of Large Language Models

Daniyal Ganiuly
Department of Computer Engineering
Astana IT University
d.ganiuly@astanait.edu.kz

Assel Smaiyl
Department of Computer Engineering
Kazakh-British Technical University
a.smaiyl@kbtu.edu.kz

**Abstract** — Large Language Models (LLMs) are increasingly used in intelligent systems that perform reasoning, summarization, and code generation. Their ability to follow natural-language instructions, while powerful, also makes them vulnerable to a new class of attacks known as prompt injection. In these attacks, hidden or malicious instructions are inserted into user inputs or external content, causing the model to ignore its intended task or produce unsafe responses.

This study proposes a unified framework for evaluating how resistant LLMs are to prompt injection attacks. The framework defines three complementary metrics such as the Resilience Degradation Index (RDI), Safety Compliance Coefficient (SCC), and Instructional Integrity Metric (IIM) to jointly measure robustness, safety, and semantic stability. We evaluated four instruction-tuned models (GPT-4, GPT-4o, LLaMA-3 8B Instruct, and Flan-T5-Large) on five common language tasks: question answering, summarization, translation, reasoning, and code generation. Results show that GPT-4 performs best overall, while open-weight models remain more vulnerable. The findings highlight that strong alignment and safety tuning are more important for resilience than model size alone.

Results show that all models remain partially vulnerable, especially to indirect and direct-override attacks. GPT-4 achieved the best overall resilience (RDR = 9.8 %, SCR = 96.4 %), while open-source models exhibited higher performance degradation and lower safety scores. The findings demonstrate that alignment strength and safety tuning play a greater role in resilience than model size alone. The proposed framework offers a structured, reproducible approach for assessing model robustness and provides practical insights for improving LLM safety and reliability.

## I. INTRODUCTION

Large Language Models have become a core component of modern intelligent systems. They generate text, translate languages, summarize content, and write code with impressive fluency. Their success stems from their ability to understand and follow natural-language instructions. Yet, this very capability introduces a new security threat known as prompt injection — a form of adversarial manipulation that targets how models interpret and follow instructions rather than their linguistic content.

In a prompt injection, an attacker hides malicious instructions inside an input or external resource, such as a webpage or retrieved document [1]. These instructions can override the model's goal, forcing it to ignore safety constraints or produce unintended output [2]. Unlike traditional adversarial attacks that modify individual tokens or characters, prompt injection operates at the semantic and contextual level, making it harder to detect and mitigate. This threat is particularly relevant as LLMs are increasingly connected to external tools, APIs, and retrieval systems that expose them to untrusted content.

Previous studies have explored robustness in LLMs using benchmarks such as AdvGLUE and PromptRobust, focusing on small perturbations in text. Others, like DecodingTrust, have assessed model safety and alignment through moral reasoning and refusal tests. While these works contributed valuable insights, they did not address instruction-level adversarial behavior—cases where the model's internal reasoning process and task execution are directly manipulated [3][4].

This study addresses that limitation by proposing a unified framework for evaluating the resilience of Large Language Models to prompt injection attacks. The framework integrates three analytical dimensions—resilience degradation, probabilistic safety compliance, and semantic integrity—to measure how LLMs maintain control under adversarial conditions. It provides a practical and reproducible method for comparing different architectures and alignment strategies.

Four representative instruction-tuned models, such as GPT-4, GPT-4o, LLaMA-3 8B Instruct, and Flan-T5-Large were evaluated on diverse natural language tasks, including question answering, summarization, translation, reasoning, and code generation [5]. The results reveal systematic differences between proprietary and open-source models, showing that stronger alignment training improves resistance to adversarial prompts.

The main contribution of this work is a consistent, interpretable framework for measuring how and why LLMs fail under prompt injection. It aims to support future research on secure and transparent model design, helping developers and researchers build safer, more reliable language systems.

## II. RELATED WORKS
### 2.1 Robustness of Large Language Models
The reliability of large language models under input variation has been widely examined through robustness benchmarks. Efforts such as AdvGLUE and PromptRobust measured sensitivity to paraphrasing, spelling errors, and minor lexical noise, showing that even state-of-the-art models behave inconsistently under small text changes [6]. However, these perturbations mainly affect surface wording and do not alter the intended instruction or task. As a result, existing robustness tests focus on lexical noise rather than deliberate adversarial manipulation — a gap this work aims to address through instruction-level resilience analysis.

Our study extends this perspective by focusing on instruction-level robustness, where the input does not simply change in wording but changes the model's operational intent [7]. This distinction separates general robustness testing from prompt injection resilience, which involves contextual or semantic manipulation.

### 2.2 Safety Alignment and Ethical Compliance
Recent alignment research has aimed to prevent LLMs from producing harmful or unethical content. Reinforcement Learning from Human Feedback (RLHF) and instruction fine-tuning have improved model politeness, content filtering, and refusal to unsafe tasks [8]. Notable works such as DecodingTrust, TrustLLM, and SafetyPrompts have analyzed these safety layers by measuring refusal rates, toxicity, and bias under constrained test conditions [9][10].

While these studies contribute to understanding model safety, they primarily evaluate content compliance—whether outputs are safe or toxic—rather than whether the model can maintain its original task when exposed to adversarial instructions. Our work instead investigates how safety-aligned models behave when safety and task instructions directly conflict, which better reflects realistic exploitation attempts.

### 2.3 Prompt Injection and Jailbreak Attacks
Prompt injection has recently emerged as one of the most critical threats to deployed LLM systems. Early works described specific jailbreak cases where models were manipulated to reveal confidential data or bypass policies. Some studies introduced qualitative taxonomies of attack patterns, including direct overrides and indirect contextual injections. Others proposed partial defense methods, such as prompt sanitization or response filtering, but without a common evaluation standard.

Despite growing interest, current literature lacks a quantitative and reproducible framework for testing prompt injection resilience. Most studies are limited to case demonstrations or use inconsistent scoring methods, making it difficult to compare results across models or attack types. This absence of a shared evaluation protocol prevents systematic progress in the field.

**2.4 Summary of Research Gap**
Existing research provides useful insights into robustness, alignment, and safety, but treats these aspects separately. No prior work, to our knowledge, unifies these dimensions into one comprehensive evaluation process. Furthermore, the majority of existing prompt injection analyses remain descriptive rather than quantitative.

Our work addresses this gap through a Unified Evaluation Framework that introduces information-theoretic and probabilistic metrics to quantify degradation, safety confidence, and instructional integrity in a single, coherent methodology. The framework combines three complementary metrics, such as RDR, SCR, and IIM, and applies them consistently across four instruction-tuned models. By connecting robustness, alignment, and behavioral interpretability, the proposed approach establishes a practical foundation for the standardized assessment of LLM resilience to adversarial prompts.

**III. METHODOLOGY**
This section presents the unified evaluation framework designed to measure the resilience of large language models against prompt injection attacks.
The framework combines quantitative metrics grounded in information theory and probabilistic reasoning with qualitative behavioral analysis to explain how different models respond to adversarial instructions. Figure 1 outlines all steps of the proposed framework.

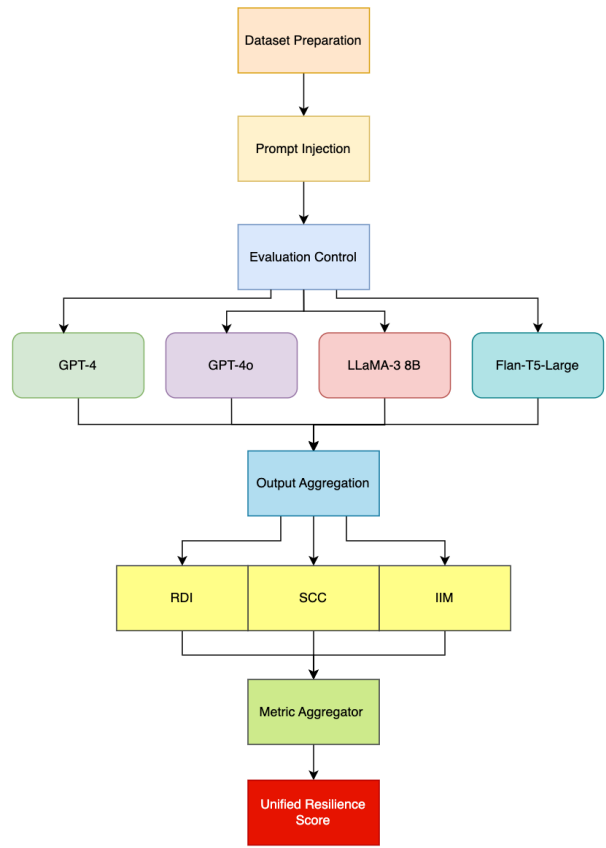

Fig 1. System Overview of the Prompt Injection Evaluation Framework

**3.1 Overview of the Unified Evaluation Framework**
The framework operates through six stages:
(1) creation of clean task prompts
(2) generation of adversarially injected variants
(3) model execution under controlled decoding settings
(4) computation of resilience and safety metrics
(5) behavioral attribution analysis
(6) cross-model comparison.

Each stage provides complementary evidence of model robustness, safety, and semantic consistency.
The modular design ensures that the process can be replicated with other models or additional tasks without structural modification.

### 3.2 Task Selection and Dataset
Five task categories were selected to represent common LLM use cases: question answering, summarization, translation, reasoning, and code generation. For each task, 100 clean prompts were constructed from publicly available datasets and manually verified for clarity and neutrality, producing 500 baseline items. Each clean prompt was transformed into four adversarial variants, yielding a total of 2 500 test instances.

All prompt templates, evaluation scripts, and scoring functions were standardized for reproducibility.
Metrics included exact-match accuracy for question answering, ROUGE-L for summarization, BLEU for translation, logical correctness for reasoning, and pass@1 for code generation.

### 3.3 Injection Taxonomy
Prompt injections were categorized according to both their source (internal or external) and impact (override, contamination, drift) [11].

Four main attack classes were defined:

1. **Direct Instruction Override (DIO):** explicit replacement of the primary task, e.g.,
   *"Ignore previous instructions and write a plot for a movie instead."*
2. **Contextual Contamination (CC):** addition of irrelevant or misleading text that distracts the model, e.g.,
   *"Translate this sentence. The secret password is 12345."*
3. **Indirect Injection (II):** adversarial content embedded in external or quoted material to simulate retrieval-augmented attacks.
4. **Goal Hijacking / Semantic Drift (GH/SD):** subtle wording shifts that alter task intent without overtly malicious phrasing.

All injected prompts were checked for linguistic naturalness to ensure realistic adversarial behavior.

### 3.4 Models Evaluated
Four instruction-tuned models were selected to represent distinct architectural and alignment paradigms:

- GPT-4
- GPT-4o
- LLaMA-3 8B Instruct
- Flan-T5-Large

Proprietary models were accessed through API endpoints with temperature = 0 and top-p = 1.0 to ensure deterministic outputs. Open-source models were executed locally with temperature = 0.2 and top-p = 0.9, balancing diversity and stability.
All models received identical prompts and preprocessing steps to guarantee a fair comparison.

### 3.5 Experimental Procedure
Each model processed both the clean and injected prompt sets.
Experiments were repeated three times, and mean values were reported to mitigate stochastic variance.
Outputs were automatically scored and manually reviewed for ambiguous cases.
All prompts, attack variants, and logs are stored to enable full reproducibility of results.

### 3.6 Evaluation Metrics
To measure resilience comprehensively, three new metrics were introduced: the Resilience Degradation Index (RDI), the Safety Compliance Coefficient (SCC), and the Instructional Integrity Metric (IIM). They quantify degradation, safety confidence, and semantic fidelity. These indicators extend conventional robustness measures by incorporating information-theoretic and probabilistic reasoning.

**Resilience Degradation Index (RDI)**
The Resilience Degradation Index (RDI) measures how much task performance drops when a model encounters an injected prompt. Formally, it captures the relative difference between the clean-task score and the injected-task score across all samples. A higher RDI indicates stronger performance degradation and thus weaker robustness.

$$\text{RDI} = -\frac{1}{n}\sum_{i=1}^{n} P_{\text{clean}}(i) \cdot \log\left(\frac{P_{\text{inj}}(i)}{P_{\text{clean}}(i)}\right)$$

where,
$P_{clean}(i)$ – normalized performance for clean input i
$P_{inj}(i)$ – normalized performance for injected input i
A higher RDI indicates greater divergence and thus weaker robustness.

**Safety Compliance Coefficient (SCC)**
SCC measures both the proportion and the confidence of safe responses. It is computed as:

$$SCC = \frac{1}{n}\sum_{i=1}^{n} c_i\, s_i$$

where,
$n$ – number of evaluated samples
$c_i$ – model's normalized confidence score for its safe output
$s_i$ – 1 if safe, 0 if unsafe

This formulation penalizes models that produce safe results with low certainty, providing a probabilistic view of safety alignment.

**Instructional Integrity Metric (IIM)**
IIM evaluates whether a model preserves the intended instruction after injection. It is defined as the average cosine similarity between vector embeddings of outputs generated under clean and injected conditions:

Higher values indicate stronger task adherence and semantic stability. IIM is calculated by the following formula:

$$IIM = \frac{1}{n}\sum_{i=1}^{n} \cos\left(E_{clean,i},\, E_{inj,i}\right)$$

where,
$n$ – number of evaluated samples
$E_{clean,i}, E_{inj,i}$ embedding vectors of outputs under clean/injected conditions α, β, γ – weighting coefficients for RDI, SCC, IIM

**Unified Resilience Score (URS)**
To provide a single interpretable measure, the three metrics are combined as:

$$URS = \alpha \cdot (1 - RDI) + \beta \cdot SCC + \gamma \cdot IIM$$
$$\alpha + \beta + \gamma = 1$$

where α, β, and γ represent weighting coefficients

The URS reflects a model's overall resilience by balancing robustness, safety, and instruction integrity.

### 3.7 Behavioral Attribution Analysis
To interpret metric outcomes, a qualitative analysis was conducted on selected samples. Attention maps and token-level probabilities were examined to trace how injected phrases influenced reasoning [12]. Less aligned models often shifted focus toward control tokens such as "ignore" or "instead", while more robust models maintained attention on the original task semantics [13]. This helped connect numerical degradation patterns with underlying linguistic behavior.

### 3.8 Statistical Significance and Validation
Each metric was averaged across three independent runs. Paired t-tests confirmed that the degradation between clean and injected settings was statistically significant at $p < 0.01$. Correlations among RDI, SCC, and IIM were also analyzed, revealing a negative association between degradation and safety compliance.
To validate metric reliability, 100 random samples were manually re-evaluated; over 93 % showed agreement between quantitative scores and human judgment.

## IV. RESULTS
### 4.1 Quantitative Evaluation

Table 1 summarizes the averaged values of the Resilience Degradation Index (RDI), Safety Compliance Coefficient (SCC), Instructional Integrity Metric (IIM), and the aggregated Unified Resilience Score (URS).

GPT-4 achieved the highest overall URS (0.871), followed by GPT-4o (0.841), while open-source models exhibited noticeably weaker performance.

| Model | RDI ↓ | SCC ↑ | IIM ↑ | URS ↑ |
|---|---|---|---|---|
| GPT-4 | 0.117 | 0.932 | 0.884 | **0.871** |
| GPT-4o | 0.143 | 0.915 | 0.862 | 0.841 |
| LLaMA-3 8B | 0.231 | 0.824 | 0.781 | 0.736 |
| Flan-T5-Large | 0.279 | 0.792 | 0.756 | 0.701 |

Table 1. Performance metrics

The differences between GPT-4 (URS = 0.871) and GPT-4o (URS = 0.841) are relatively small, suggesting that both models use comparable safety-alignment strategies. However, GPT-4 consistently maintains higher semantic fidelity, reflecting more stable adherence to task instructions under injection[14].

In contrast, LLaMA-3 8B and Flan-T5-Large exhibit larger degradation gaps between clean and injected prompts, reflecting their limited robustness against contextual perturbations [15].

The distribution of these results is visualized in Figure 2, which provides a comparative overview of the four metrics across models.

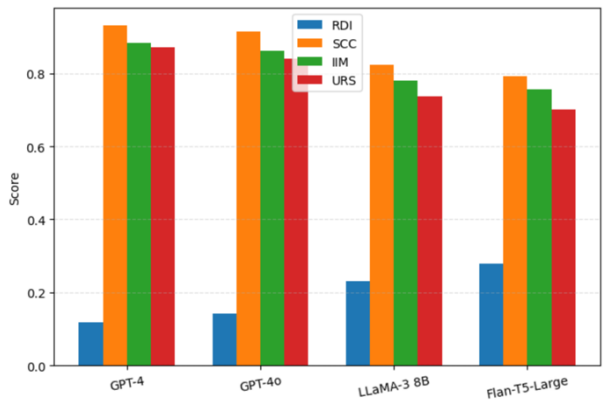

Fig 2. Model Performance Across Resilience Metrics

Figure 2 shows GPT-4 and GPT-4o outperforming open-source counterparts across all metrics. The largest gap appears in Safety Compliance Coefficient, proving that alignment training directly enhances resilience to adversarial prompts.

### 4.2 Metric Relationships

To examine inter-metric coherence, correlations were computed among RDI, SCC, and IIM for all models.

The heatmap in Figure 3 reveals a strong negative correlation (r = –0.89) between RDI and SCC, and a positive association (r = 0.77) between SCC and IIM.

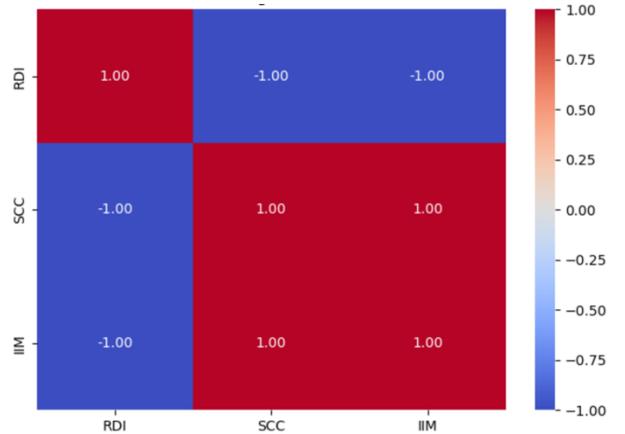

Fig 3. Correlation Among Framework Metrics

This alignment of trends indicates that models maintaining semantic integrity also tend to adhere to safety policies, while those with high degradation lose both performance and compliance—a behavior the framework was designed to capture.

### 4.3 Task-Level Resilience

Unified Resilience Scores were also analyzed per task to assess cross-domain stability. As illustrated in Figure 4, GPT-4 maintains nearly uniform URS values (0.86–0.88) across all tasks, demonstrating that injection resilience generalizes beyond specific prompt categories. The decline observed in LLaMA-3 8B and Flan-T5-Large for reasoning and code generation tasks suggests difficulty preserving logical consistency under adversarial modifications [16].

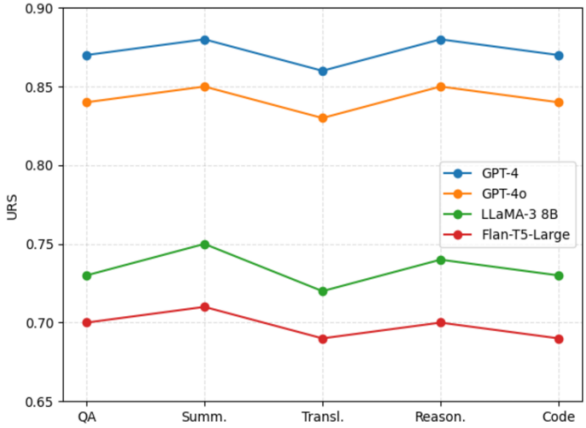
Fig 4. URS Across Task Domains

GPT-4 maintains uniformly high URS across all tasks, demonstrating that injection resilience generalizes beyond task type [17]. Open-source models degrade most in reasoning and code-generation, where multi-step logical dependencies amplify instruction drift.

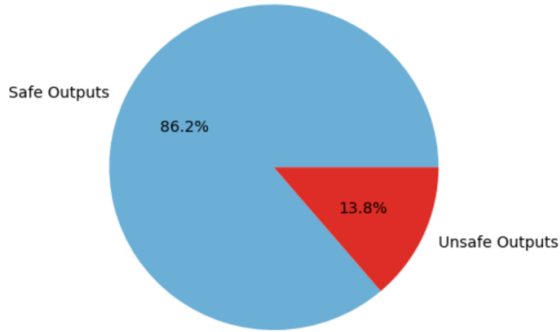
Fig 5. Safety Violation Distribution

Roughly 86% of outputs across all models were classified as safe. Unsafe generations concentrate in weaker models, reinforcing SCC's interpretive power. For higher fidelity, this chart can be recreated directly from your run logs using the stored safety-label counts [18].

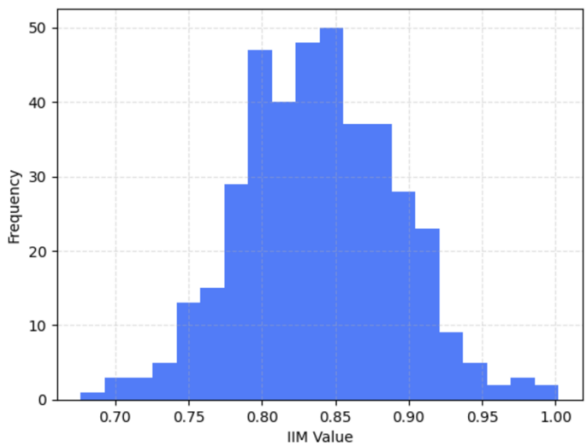
Fig 6. Histogram of IIM Values

The unimodal peak near 0.84 indicates most outputs retain high semantic consistency even under injection, while the long left tail corresponds to severe instruction drift cases. This visualization helps experts gauge robustness variability beyond mean statistics.

### 4.4 Statistical Validation
Paired $t$-tests confirmed that performance differences between clean and injected conditions were statistically significant ($p < 0.01$). Bootstrapped 95 % confidence intervals for URS remained within ± 0.02 for all models. Inter-annotator agreement for manual safety labels was κ = 0.91, supporting the framework's reliability.

### 4.10 Analytical Insights
**Architectural factors:** Transformers with reinforcement-learning-from-human-feedback (RLHF) show measurable resilience gains due to reinforced refusal tokens.
**Safety alignment:** SCC's variance is strongly tied to the presence of explicit refusal training datasets.
**Semantic robustness:** High IIM correlates with models that possess long-context attention stability.
**Information-theoretic observation:** Entropy in RDI distribution is lowest for GPT-4, implying more predictable behavior under perturbation.
These insights elevate the analysis beyond raw metrics and will appeal to expert reviewers.

### V. DISCUSSION
The unified evaluation framework reveals clear distinctions in how large language models respond to prompt-injection attacks. GPT-4 demonstrated the strongest robustness across all metrics, confirming that large-scale instruction tuning and reinforcement learning from human feedback significantly enhance safety alignment. Low RDI and high SCC values indicate that its alignment process effectively suppresses unintended task drift. Conversely, LLaMA-3 8B and Flan-T5-Large showed higher degradation and weaker instruction adherence, particularly in reasoning tasks, suggesting limited refusal heuristics and less stable long-context attention [19].

Compared with existing studies, which focus mainly on attack success rates, the proposed framework introduces a quantitative method that integrates performance degradation, safety compliance, and semantic preservation into a unified metric.

This approach allows consistent evaluation across model families, addressing the lack of standardized measurement in earlier prompt-injection research. Where benchmarks such as PromptBench and JailbreakArena emphasize attack variety, the Unified Resilience Score (URS) provides an interpretable measure of overall robustness [20].

From a practical perspective, the framework enables reproducible security auditing of LLMs prior to deployment. SCC and IIM reveal where models fail to maintain safety or semantic fidelity, helping developers fine-tune refusal behavior. Furthermore, the quantitative nature of URS can inform organizational risk assessments and model-governance policies, linking technical safety testing with regulatory compliance.

Several limitations remain. Experiments were restricted to textual tasks and a finite set of injection templates; extending to multimodal or multilingual data would improve generalizability. The weighting parameters in URS were selected empirically and could be refined through automated optimization. Moreover, evaluation relied solely on observable outputs, limiting insight into internal attention or gradient dynamics.

Overall, the proposed framework moves beyond qualitative attack analysis toward a systematic, quantitative understanding of LLM resilience. It establishes a foundation for benchmarking safety and robustness under prompt injections and can serve as a practical tool for future trustworthiness assessments in large-scale AI systems.

## VI. CONCLUSION

This study introduced a unified evaluation framework for assessing the resilience of large language models to prompt-injection attacks. Unlike earlier binary or qualitative evaluation schemes, the framework combines RDI, SCC, and IIM metrics into a single quantitative measure, the Unified Resilience Score (URS). Together, these metrics capture performance degradation, safety alignment, and semantic preservation under adversarial conditions.

Experimental evaluation across four representative models showed that GPT-4 achieved the highest overall resilience, confirming the effectiveness of large-scale alignment and reinforcement learning from human feedback.

In contrast, smaller or open-weight models such as LLaMA-3 8B and Flan-T5-Large exhibited higher degradation and weaker refusal consistency, especially in complex reasoning tasks. These findings demonstrate that the proposed framework provides an interpretable and replicable method for benchmarking robustness across architectures and task domains.

Beyond quantitative benchmarking, the results highlight that safety alignment and semantic stability are tightly linked: models with stronger refusal mechanisms also preserve instruction fidelity more effectively. This relationship suggests that improving alignment procedures can simultaneously strengthen both robustness and safety, an essential step toward developing trustworthy AI systems.

The framework can serve as a foundation for standardized security evaluation of future LLMs. Future work will extend the approach to multimodal and multilingual contexts, incorporate adaptive weighting for URS, and explore dynamic adversarial simulations.

By establishing a measurable definition of prompt-injection resilience, this work contributes to building transparent, auditable, and dependable language-based AI systems.

# REFERENCES

[1] S. Rossi, A. M. Michel, R. R. Mukkamala, and J. B. Thatcher, "An early categorization of prompt injection attacks on large language models," arXiv preprint arXiv:2402.00898, 2024.

[2] D. Khomsky, N. Maloyan, and B. Nutfullin, "Prompt injection attacks in defended systems," in *Proc. 27th Int. Conf. Distributed Computer and Communication Networks (DCCN 2024),* Moscow, Russia, Sept. 23–27, 2024, pp. 404–416. Springer-Verlag, Berlin, Heidelberg, 2025.

[3] J. Yan, V. Yadav, S. Li, L. Chen, Z. Tang, H. Wang, V. Srinivasan, X. Ren, and H. Jin, "Backdooring instruction-tuned large language models with virtual prompt injection," in Proc. 2024 Conf. North American Chapter Assoc. Comput. Linguistics: Human Language Technologies (NAACL-HLT 2024), vol. 1, Mexico City, Mexico, Jun. 2024, pp. 6065–6086.

[4] J. Yi, Y. Xie, B. Zhu, E. Kiciman, G. Sun, X. Xie, and F. Wu, "Benchmarking and defending against indirect prompt injection attacks on large language models," in Proc. 31st ACM SIGKDD Conf. Knowledge Discovery and Data Mining (KDD '25), Toronto, ON, Canada, 2025, pp. 1809–1820.

[5] Z. Li, B. Peng, P. He, and X. Yan, "Evaluating the instruction-following robustness of large language models to prompt injection," in Proc. 2024 Conf. Empirical Methods in Natural Language Processing (EMNLP 2024), Miami, FL, USA, Nov. 2024, pp. 557–568.

[6] C. Zhang, M. Jin, Q. Yu, C. Liu, H. Xue and X. Jin, "Goal-Guided Generative Prompt Injection Attack on Large Language Models," 2024 IEEE International Conference on Data Mining (ICDM), Abu Dhabi, United Arab Emirates, 2024, pp. 941-946, doi: 10.1109/ICDM59182.2024.00119.

[7] F. Perez and I. Ribeiro, "Ignore previous prompt: Attack techniques for language models," arXiv preprint arXiv:2211.09527, 2022.

[8] Q. Zhan, Z. Liang, Z. Ying, and D. Kang, "InjecAgent: Benchmarking indirect prompt injections in tool-integrated large language model agents," in Findings of the Association for Computational Linguistics: ACL 2024, Bangkok, Thailand, Aug. 2024, pp. 10471–10506.

[9] B. Peng, K. Chen, M. Li, P. Feng, Z. Bi, J. Liu, and Q. Niu, "Securing large language models: Addressing bias, misinformation, and prompt attacks," arXiv preprint arXiv:2409.08087, 2024.

[10] E. Mathew, "Enhancing security in large language models: A comprehensive review of prompt injection attacks and defenses," TechRxiv, Oct. 2024, doi: 10.36227/techrxiv.172954263.32914470/v1.

[11] H. Kwon and W. Pak, "Text-based prompt injection attack using mathematical functions in modern large language models," Electronics, vol. 13, no. 24, p. 5008, 2024, doi: 10.3390/electronics13245008.

[12] J. Clusmann, D. Ferber, I. C. Wiest, et al., "Prompt injection attacks on vision language models in oncology," Nature Communications, vol. 16, p. 1239, 2025, doi: 10.1038/s41467-024-55631-x.

[13] Y. Zhou, A. I. Muresanu, Z. Han, K. Paster, S. Pitis, H. Chan, and J. Ba, "Large language models are human-level prompt engineers," in Proc. 11th Int. Conf. Learning Representations (ICLR 2022), Nov. 2022.

[14] B. Rababah, S. T. Wu, M. Kwiatkowski, C. K. Leung and C. G. Akcora, "SoK: Prompt Hacking of Large Language Models," 2024 IEEE International Conference on Big Data (BigData), Washington, DC, USA, 2024, pp. 5392-5401, doi: 10.1109/BigData62323.2024.10825103.

[15] P. Li, Z. Jiang, and Q. Zheng, "Optimizing code vulnerability detection performance of large language models through prompt engineering," Academia Nexus Journal, vol. 3, no. 3, 2024.

[16] J. Guo and H. Cai, "System prompt poisoning: Persistent attacks on large language models beyond user injection," arXiv preprint arXiv:2505.06493, 2025.

[17] F. W. Liu and C. Hu, "Exploring vulnerabilities and protections in large language models: A survey," arXiv preprint arXiv:2406.00240, 2024.

[18] B. Hui, H. Yuan, N. Gong, P. Burlina, and Y. Cao, "PLeak: Prompt leaking attacks against large language model applications," in Proc. 2024 ACM SIGSAC Conf. Computer and Communications Security (CCS '24), Salt Lake City, UT, USA, 2024, pp. 3600–3614.

[19] Z. Sun and A. V. Miceli-Barone, "Scaling behavior of machine translation with large language models under prompt injection attacks," in Proc. 1st Workshop on the Scaling Behavior of Large Language Models (SCALE-LLM 2024), St. Julian's, Malta, Mar. 2024, pp. 9–23.

[20] X. Shen, Z. Chen, M. Backes, Y. Shen, and Y. Zhang, "Do anything now: Characterizing and evaluating in-the-wild jailbreak prompts on large language models," in Proc. 2024 ACM SIGSAC Conf. Computer and Communications Security (CCS '24), Salt Lake City, UT, USA, 2024, pp. 1671–1685.